# Colloidal aggregation and critical Casimir forces - Reply

In our Letter [1], we presented experiments on critical Casimir induced colloidal aggregation in a system with negligibly small van der Waals forces. The latter are usually inducing aggregation in colloidal systems, but by refractive-index matching, we were able to minimize these forces. Aggregation was nonetheless observed; in order to interpret the observed aggregation, we proposed that this was due to a competition between repulsive electrostatic and attractive Casimir forces. We presented a simple model based on a competition between these two forces, that was able to account in a rather satisfactory way for the experimental data, without any adjustable parameters. Gambassi and Dietrich [2] propose that the simple model should be regarded with caution as to its domain of application, and that it can be refined in a number of different ways.

As far as the refinements of our model are concerned, indeed the line $l_d = \xi$ illustrates where the repulsive and attractive components of the potential have equal range, and this does not exactly coincide with where aggregation is expected to occur, since for that a ~3kT minimum has to develop in the potential. This was mentioned explicitly in our original Letter [1]. However, as was also mentioned, the minimum develops very rapidly when changing the temperature, so that $l_d = \xi$ is a reasonable approximation. From the results presented in [2], this approximation does, in fact, seem to work very well for most of our data points as shown in [2], except for the last data point, for which both the Debye length and the bulk correlation lengths are large.

As far as the domain of applicability of the model is concerned, indeed critical scaling of, for instance, the bulk correlation length is supposed to hold only very close to the critical point, which is not necessarily the case for all our data points. This is of course true, and this is the reason why we determined the bulk correlation length using independent measurements. However, it is in general not clear how close to the critical point one should be, as in experiments critical scaling is often found to still correctly describe the behavior rather far from the critical point. Notably, for critical adsorption, a situation akin to our experiments (cite Law), the proposed critical scaling laws remain verified at large distances from the critical point; e.g. in [3] theory and experiment agree even for a reduced temperature distance from the critical point $t = \frac{|T_c - T|}{T_c} \approx 0.1$, corresponding to a very small bulk correlation length.

Both issues – the actual form of the potential and the possible absence of critical scaling- could in principle be resolved by a complete calculation of the Casimir interaction without any approximations which appears to be feasible [4,5]. Until this is done, it seems difficult to assess e.g. the reason for the deviation of the last data point from our simple arguments in [1] improved in [2],

Daniel Bonn[1,2], Gerard Wegdam[1] and Peter Schall[1]
[1]van der Waals-Zeeman Institute, University of Amsterdam, Valckenierstraat 65, NL-1018XE Amsterdam
[2]Laboratoire de Physique Statistique de l'ENS, 24 Rue Lhomond, F-75231 Paris cedex 05